\documentclass[
  aps,
  prb,
  reprint,
  superscriptaddress,
  nofootinbib,
  raggedbottom,
  floatfix,
  longbibliography
]{revtex4-2}

\usepackage{xcolor}
\definecolor{guanpurple}{RGB}{122,52,174}

\usepackage{orcidlink}

\usepackage{amsmath}
\usepackage{amssymb}
\usepackage{bm}
\usepackage{graphicx}
\usepackage{dcolumn}
\usepackage{hyperref}


\begin{document}

\title{
Role of crystal-field degrees of freedom in square and rectangular skyrmion lattices in noncentrosymmetric tetragonal magnets
}

\author{Cong~Guan}
\affiliation{
Graduate School of Science, Hokkaido University,
Sapporo 060-0810, Japan
}

\author{Yan~S.~Zha~\orcidlink{0009-0008-8568-446X}}
\affiliation{
Graduate School of Science, Hokkaido University,
Sapporo 060-0810, Japan
}

\author{Satoru~Hayami~\orcidlink{0000-0001-9186-6958}}
\affiliation{
Graduate School of Science, Hokkaido University,
Sapporo 060-0810, Japan
}
\begin{abstract}
We investigate the role of crystal-field degrees of freedom in stabilizing square and rectangular skyrmion lattices in a noncentrosymmetric localized-moment model on a square lattice.
The magnetic phases are determined using self-consistent mean-field calculations.
We first consider a single crystal-field doublet with exchange and polar-type Dzyaloshinskii--Moriya interactions.
We find square skyrmion lattices with distinct magnetic periods, as well as a rectangular skyrmion lattice that spontaneously breaks the in-plane fourfold rotational symmetry.
Their stability and field-induced phase sequences are strongly controlled by the crystal-field wave function, including reentrant behavior of the rectangular phase.
Second-neighbor exchange further modifies the competition among these states.
When a second crystal-field doublet is retained, additional
rectangular skyrmion lattices with distinct modulation patterns
emerge, together with axial single-$Q$ states and
various double-$Q$ states.
These results demonstrate that crystal-field degrees of freedom provide a microscopic mechanism for tuning the stability, spatial symmetry, and topology of multiple-$Q$ magnetic states.
\end{abstract}
\maketitle
\section{Introduction}

Magnetic skyrmions are noncoplanar spin textures characterized by a nonzero topological charge~\cite{Bogdanov89, Bogdanov94} and have attracted considerable interest owing to their distinctive magnetic and transport properties~\cite{nagaosa2013topological,fert2017magnetic, bogdanov2020physical, Tokura_doi:10.1021/acs.chemrev.0c00297, hayami2024stabilization}.
A well-established mechanism for stabilizing skyrmions is the Dzyaloshinskii--Moriya interaction (DMI), which originates from spin--orbit coupling in systems with broken inversion symmetry~\cite{dzyaloshinsky1958thermodynamic,moriya1960anisotropic}.
The competition between exchange interactions, DMI, magnetic anisotropy, and an external magnetic field can stabilize periodic skyrmion lattices (SkLs)~\cite{rossler2006spontaneous, Rowland_PhysRevB.93.020404}.
Experimentally, SkLs have been observed in various noncentrosymmetric magnets. 
In chiral magnets such as MnSi~\cite{Muhlbauer_2009skyrmion, Neubauer_PhysRevLett.102.186602, Adams_PhysRevLett.107.217206}, Fe$_{1-x}$Co$_{x}$Si~\cite{yu2010real, adams2010skyrmion, Munzer_PhysRevB.81.041203, Bauer_PhysRevB.93.235144}, and EuPtSi~\cite{kakihana2018giant,kaneko2019unique,tabata2019magnetic, kakihana2019unique, hayami2021field, Matsumura_PhysRevB.109.174437}, the DMI favors Bloch-type skyrmions. 
By contrast, N\'eel-type SkLs have been observed in 
polar magnets such as GaV$_4$S$_8$~\cite{kezsmarki_neel-type_2015, ruff2015multiferroicity, Zhang_PhysRevB.95.085136, Gross_PhysRevB.102.104407} and VOSe$_2$O$_5$~\cite{Kurumaji_PhysRevLett.119.237201, kurumaji2021direct, Kuo2025enhanced}, while a hybrid-type SkL has been reported in the polar magnet EuNiGe$_3$~\cite{singh2023transition, matsumura2024helicity}.
These observations demonstrate that crystal symmetry and the associated form of the DMI play an important role in determining the internal structure of skyrmions.

The discovery of SkLs and related multiple-$Q$ magnetic textures in centrosymmetric rare-earth intermetallics has broadened the microscopic mechanisms for topological and modulated magnetism.
Examples include Gd$_2$PdSi$_3$~\cite{Saha_PhysRevB.60.12162, kurumaji2019skyrmion, sampathkumaran2019report, Spachmann_PhysRevB.103.184424, Gomilsek_PhysRevLett.134.046702}, GdRu$_2$Si$_2$~\cite{khanh2020nanometric, Wood_PhysRevB.107.L180402, eremeev2023insight, Huddart_PhysRevB.111.054440, Spethmann_PhysRevMaterials.8.064404, dong2025pseudogap}, and EuAg$_4$Sb$_2$~\cite{neves2026cascade, neves2026general}, where competing interactions give rise to a variety of field-induced multiple-$Q$ and long-period magnetic states.
In centrosymmetric metallic magnets, interactions mediated by itinerant electrons can promote multiple-$Q$ instabilities and stabilize SkLs even in the absence of DMI~\cite{Akagi_PhysRevLett.108.096401, Hayami_PhysRevB.95.224424, Wang_PhysRevLett.124.207201}.
In particular, the square SkL observed in GdRu$_2$Si$_2$ has been discussed theoretically in connection with various microscopic mechanisms that stabilize multiple-$Q$ states with fourfold symmetry~\cite{Yasui2020imaging, khanh2022zoology}, such as anisotropic exchange interactions, effective biquadratic interactions arising from itinerant electrons, and single-ion anisotropy~\cite{Hayami_doi:10.7566/JPSJ.89.103702, Hayami_PhysRevB.103.024439,
Utesov_PhysRevB.103.064414, Wang_PhysRevB.103.104408,
Bouaziz_PhysRevLett.128.157206, nomoto2023ab,
utesov2025thermodynamic, wood2025magnon}.
Such studies have shown that anisotropic interactions and interactions at higher-harmonic wave vectors can play an important role in selecting square and rectangular magnetic textures \cite{Hayami_PhysRevB.105.174437, hayami2023widely}.

In rare-earth magnets, the local crystal-field states provide another important ingredient. The magnetic matrix elements of localized $4f$ electrons depend on the crystal-field wave functions and can therefore
strongly affect the magnetic anisotropy and effective interactions~\cite{stevens1952matrix, hutchings1964point, Jensen1991, Christianson_PhysRevB.66.193102, Willers_PhysRevB.81.195114, willers2015correlation}. 
Recent mean-field studies of centrosymmetric tetragonal systems have shown that multiple low-lying Kramers doublets can stabilize square SkLs and substantially modify their magnetic structures~\cite{Zha_PhysRevB.111.165155, zha2026effects}.
Unlike a phenomenological magnetic anisotropy introduced as an independent coupling, the crystal-field wave function simultaneously determines the matrix elements entering the exchange, DMI, and Zeeman terms.

This raises the question of how such crystal-field effects influence skyrmion formation in noncentrosymmetric $f$-electron magnets, where the DMI provides an additional mechanism for stabilizing modulated magnetic states.
In particular, it remains unclear how the wave-function composition of a Kramers doublet affects the stability of DMI-induced SkLs and their spontaneous breaking of fourfold rotational symmetry, or how these magnetic states evolve when additional crystal-field levels become relevant.

In this paper, we investigate these issues using a localized $4f$-electron model on a two-dimensional square lattice with exchange interactions, polar-type DMI, and a magnetic field perpendicular to the plane. 
We consider two tetragonal $\Gamma_7$ Kramers doublets and examine the evolution from the single-doublet limit to the regime where both doublets participate in the magnetic states. 
Using self-consistent mean-field calculations, we find square SkLs with distinct magnetic periods and a rectangular SkL that spontaneously breaks the in-plane fourfold rotational symmetry.
The stability of these states is strongly affected by the crystal-field wave function.
We further examine the effects of second-neighbor exchange and show that the inclusion of the second doublet gives rise to additional multiple-$Q$ states and rectangular SkLs with distinct Fourier contents and topological charges.
These results demonstrate that crystal-field wave functions and multiorbital degrees of freedom provide a microscopic means of controlling the stability, spatial symmetry, and topology of DMI-induced SkLs in noncentrosymmetric $f$-electron systems.

The remainder of this paper is organized as follows.
In Sec.~\ref{sec:model}, we introduce the localized $4f$-electron model, including the crystal-field basis, exchange and DMI interactions, and the single- and two-orbital formulations.
In Sec.~\ref{sec:method}, we describe the self-consistent mean-field method and the quantities used to characterize the magnetic states.
In Sec.~\ref{sec:results}, we present the magnetic phase diagrams, first for the single-orbital model and then for its extensions including second-neighbor exchange and a second crystal-field doublet.
Finally, Sec.~\ref{sec:conclusion} summarizes the main results.

\section{Model}
\label{sec:model}

We consider localized $4f$ electrons on a two-dimensional square lattice with exchange and polar-type DMI.
We first formulate a two-orbital model based on two tetragonal Kramers doublets and then obtain the single-orbital model by projection onto the lowest doublet.
The single-orbital model with nearest-neighbor exchange is used as the main model, while second-neighbor exchange and the second crystal-field doublet are introduced as extensions.
We set the lattice constant $a=1$ and use $\hbar=k_{\mathrm B}=1$ throughout.

\subsection{Local crystal-field basis}

With Ce$^{3+}$ ions in mind, we assume an $f^1$ configuration at each lattice site.
Atomic spin--orbit coupling gives the $J=5/2$ ground-state multiplet~\cite{Yosida1996}, which is split into three Kramers doublets by a tetragonal crystalline electric field.
Among them, we retain the two $\Gamma_7$ doublets and neglect the remaining doublet.
Their wave functions are
\begin{equation}
\begin{aligned}
\left|\Gamma_{7\pm}^{t(1)}\right\rangle
 &=\alpha|5/2,\pm5/2\rangle-\beta|5/2,\mp3/2\rangle,\\
\left|\Gamma_{7\pm}^{t(2)}\right\rangle
 &=\beta|5/2,\pm5/2\rangle+\alpha|5/2,\mp3/2\rangle,
\end{aligned}
\label{model:cef_basis}
\end{equation}
where $\alpha$ and $\beta=\sqrt{1-\alpha^2}$ are real coefficients.
Hereafter, we refer to each Kramers doublet as an orbital.  For convenience, we denote
$|\Gamma_{7+}^{t(1)}\rangle=|1\rangle$,
$|\Gamma_{7-}^{t(1)}\rangle=|2\rangle$,
$|\Gamma_{7+}^{t(2)}\rangle=|3\rangle$,
and $|\Gamma_{7-}^{t(2)}\rangle=|4\rangle$.
In this basis, the total-angular-momentum operators have the following matrix
representations: 
\begin{widetext}
\begin{equation}
J^x=\begin{pmatrix}
0 & -\sqrt5\alpha\beta & 0 & \frac{\sqrt5}{2}(\alpha^2-\beta^2)\\
-\sqrt5\alpha\beta & 0 & \frac{\sqrt5}{2}(\alpha^2-\beta^2) & 0\\
0 & \frac{\sqrt5}{2}(\alpha^2-\beta^2) & 0 & \sqrt5\alpha\beta\\
\frac{\sqrt5}{2}(\alpha^2-\beta^2) & 0 & \sqrt5\alpha\beta & 0
\end{pmatrix},
\label{model:Jx}
\end{equation}
\begin{equation}
J^y=\begin{pmatrix}
0 & i\sqrt5\alpha\beta & 0 & -\frac{i\sqrt5}{2}(\alpha^2-\beta^2)\\
-i\sqrt5\alpha\beta & 0 & \frac{i\sqrt5}{2}(\alpha^2-\beta^2) & 0\\
0 & -\frac{i\sqrt5}{2}(\alpha^2-\beta^2) & 0 & -i\sqrt5\alpha\beta\\
\frac{i\sqrt5}{2}(\alpha^2-\beta^2) & 0 & i\sqrt5\alpha\beta & 0
\end{pmatrix},
\label{model:Jy}
\end{equation}
\begin{equation}
J^z=\begin{pmatrix}
\frac12(5\alpha^2-3\beta^2) & 0 & 4\alpha\beta & 0\\
0 & \frac12(3\beta^2-5\alpha^2) & 0 & -4\alpha\beta\\
4\alpha\beta & 0 & \frac12(5\beta^2-3\alpha^2) & 0\\
0 & -4\alpha\beta & 0 & \frac12(3\alpha^2-5\beta^2)
\end{pmatrix}.
\label{model:Jz}
\end{equation}
\end{widetext}

\subsection{Model Hamiltonian}

The Hamiltonian is given by
\begin{equation}
H=H_{\mathrm{ex}}+H_{\mathrm{DM}}+H_\Delta+H_{\mathrm Z}.
\label{model:Htot}
\end{equation}
We consider the bilinear exchange interaction
\begin{equation}
H_{\mathrm{ex}}
=-\sum_{n=1}^{2}J_n
\sum_{\langle i,j\rangle_n}
\mathbf J_i\cdot\mathbf J_j,
\label{model:Hex}
\end{equation}
where $J_1$ and $J_2$ denote the nearest- and second-neighbor exchange interactions, respectively.
We take $J_1=1$ as the energy unit.
The second-neighbor interaction $J_2$ is introduced later as an extension of the nearest-neighbor model.
By introducing additional exchange frustration, $J_2$ modifies the competition among different ordering wave vectors and is therefore expected to affect the stability of the SkLs, as found previously in centrosymmetric magnets~\cite{Okubo_PhysRevLett.108.017206, leonov2015multiply, hayami2020multiple, Hayami_PhysRevB.94.174420, Hayami_PhysRevB.103.224418}.

Broken inversion symmetry along the $z$ direction allows a polar-type DMI,
\begin{equation}
H_{\mathrm{DM}}
=-\sum_{\langle i,j\rangle_1}
\mathbf D_{ij}\cdot
(\mathbf J_i\times\mathbf J_j),
\label{model:HDM}
\end{equation}
where
\begin{equation}
\mathbf D_{i,i+\boldsymbol\delta}
=D\,\hat{\mathbf z}\times\boldsymbol\delta
\qquad
(\boldsymbol\delta=\hat{\mathbf x},\hat{\mathbf y}).
\label{model:Dvector}
\end{equation}
This DMI favors N\'eel-type rotations of the magnetic moments.
Symmetry also allows symmetric anisotropic exchange interactions~\cite{kaplan1983single, Maksimov_PhysRevX.9.021017, Matsumoto_PhysRevB.101.224419}, which can modify the stability of multiple-$Q$ states~\cite{amoroso2020spontaneous, amoroso2021tuning, Hayami_PhysRevB.105.104428, Yambe_PhysRevB.106.174437}.
We neglect these interactions here to isolate the roles of the polar-type DMI and the crystal-field degrees of freedom.

The energy splitting between the two $\Gamma_7$ doublets is introduced as
\begin{equation}
H_\Delta=\Delta\sum_i\bigl(|3\rangle_i\,{}_i\langle3|+|4\rangle_i\,{}_i\langle4|\bigr), 
\label{model:HDelta}
\end{equation}
where the energy of the first doublet is set to zero.
In the limit $\Delta\to\infty$, the model reduces to the first-doublet model discussed below.

Finally, a magnetic field perpendicular to the lattice plane is introduced through
\begin{equation}
H_{\mathrm Z}=-h\sum_iJ_i^z,
\label{model:HZ}
\end{equation}
where $h$ denotes the Zeeman energy and we renormalize the Land\'e $g$ factor into $h$ for simplicity.

\subsection{Single-orbital model}

The single-orbital model is obtained by projecting the Hamiltonian onto the first $\Gamma_7$ doublet.
The projected angular-momentum operators are
\begin{equation}
\begin{aligned}
J_{(1)}^x&=-\sqrt5\alpha\beta\,\sigma^x,\\
J_{(1)}^y&=-\sqrt5\alpha\beta\,\sigma^y,\\
J_{(1)}^z&=\frac{5\alpha^2-3\beta^2}{2}\,\sigma^z,
\end{aligned}
\label{model:projection}
\end{equation}
where $\sigma^\mu$ are the $2\times 2$ Pauli matrices in the $\Gamma_7$-doublet space.
Thus, $\alpha$ controls the relative magnitudes of the in-plane and out-of-plane magnetic moments and consequently the effective magnetic anisotropy.

Then, the projected Hamiltonian is
\begin{equation}
H_{\mathrm{single}}
=H_{\mathrm{ex}}^{(1)}
+H_{\mathrm{DM}}^{(1)}
+H_{\mathrm Z}^{(1)},
\label{model:Hsingle}
\end{equation}
where $\mathbf J_i$ in Eqs.~\eqref{model:Hex}, \eqref{model:HDM}, and \eqref{model:HZ} is replaced by its projection $\mathbf J_{i,(1)}$ onto the $\Gamma_7$-doublet space.
We mainly study this model for $J_2=0$ as a function of $\alpha$ and $h$. 
Importantly, varying $\alpha$ does not merely introduce an independent magnetic anisotropy. 
It simultaneously changes the projected matrix elements entering the exchange, DMI, and Zeeman couplings, thereby modifying their effective balance within the low-energy doublet.

\subsection{Parameter choice for the model}
\label{model:parameter_choice}

We choose the interaction parameters such that the characteristic diagonal wave vectors are fixed at
\begin{equation}
\mathbf Q_1=(Q,Q),\qquad
\mathbf Q_2=(Q,-Q),
\qquad Q=\frac{\pi}{3},
\label{model:Qpair}
\end{equation}
which are compatible with the $6\times6$
periodic computational cell
used in the mean-field calculations.
For later use, we also define the combination wave vectors
\begin{equation}
\mathbf Q_3=(2Q,0),\qquad
\mathbf Q_4=(0,2Q),
\end{equation}
and the axial wave vectors
\begin{equation}
\mathbf Q_5=(Q,0),\qquad
\mathbf Q_6=(0,Q).
\end{equation}

To keep the wave vectors $\mathbf Q_1$ and $\mathbf Q_2$ stationary upon varying $J_2$, we choose the DMI strength as
\begin{equation}
|D|=\sqrt6\,(J_1+J_2).
\label{model:Dpath}
\end{equation}
The largest eigenvalue of the exchange--DMI interaction matrix at each wave vector is
\begin{equation}
\begin{aligned}
\lambda(\mathbf q)
&=2J_1(\cos q_x+\cos q_y)+4J_2\cos q_x\cos q_y\\
&\quad+2|D|\sqrt{\sin^2q_x+\sin^2q_y},
\end{aligned}
\label{model:lambda}
\end{equation}
where $\mathbf q=(q_x,q_y)$ is the wave vector in the first Brillouin zone.
With the choice in Eq.~\eqref{model:Dpath}, $\mathbf Q_1$ and $\mathbf Q_2$ with $Q=\pi/3$ are stationary points of $\lambda(\mathbf q)$. 
Further details of the momentum-space interactions and
the parameter path used for varying $J_2$ are provided in
Appendices~\ref{app:fourier_interaction}
and~\ref{app:J2_parameter_path}, respectively.

\section{Method}
\label{sec:method}

\subsection{Self-consistent mean-field calculations}

We determine the magnetic states using a real-space
self-consistent mean-field approximation following
Refs.~\cite{Weiss1907, Zha_PhysRevB.111.165155, zha2026effects}.
Introducing the local mean moments
$\mathbf m_i=\langle\mathbf J_i\rangle$, the exchange and DMI terms are
decoupled as
\begin{equation}
\begin{aligned}
H_{\mathrm{ex}}^{\mathrm{MF}}
={}&-\sum_{n=1}^{2}J_n\sum_{\langle i,j\rangle_n}
\left(
\mathbf J_i\cdot\mathbf m_j
+\mathbf m_i\cdot\mathbf J_j
-\mathbf m_i\cdot\mathbf m_j
\right),
\end{aligned}
\label{method:HexMF}
\end{equation}
and
\begin{equation}
\begin{aligned}
H_{\mathrm{DM}}^{\mathrm{MF}}
={}&-\sum_{\langle i,j\rangle_1}\mathbf D_{ij}\cdot
\left(
\mathbf J_i\times\mathbf m_j
+\mathbf m_i\times\mathbf J_j
-\mathbf m_i\times\mathbf m_j
\right).
\end{aligned}
\label{method:HDMMF}
\end{equation}
The crystal-field and Zeeman terms are kept unchanged.
For the single-orbital model, $\mathbf J_i$ is replaced by its projected
form $\mathbf J_{i,(1)}$ and $H_\Delta$ is omitted.

For a given set of mean moments, the mean-field Hamiltonian can be written as
\begin{equation}
H^{\mathrm{MF}}=\sum_i H_i^{\mathrm{loc}}+E_{\mathrm{dc}},
\label{method:localH}
\end{equation}
where the double-counting correction is
\begin{equation}
\begin{aligned}
E_{\mathrm{dc}}
=\sum_{n=1}^{2}J_n\sum_{\langle i,j\rangle_n}
\mathbf m_i\cdot\mathbf m_j+\sum_{\langle i,j\rangle_1}
\mathbf D_{ij}\cdot
(\mathbf m_i\times\mathbf m_j).
\end{aligned}
\label{method:Edc}
\end{equation}
We diagonalize each local Hamiltonian,
\begin{equation}
H_i^{\mathrm{loc}}|i\ell\rangle
=\epsilon_{i\ell}|i\ell\rangle,
\label{method:diagonalization}
\end{equation}
and evaluate a local observable $O_i$ as
\begin{equation}
\langle O_i\rangle
=
\frac{1}{Z_i}
\sum_\ell
e^{-\epsilon_{i\ell}/T}
\langle i\ell|O_i|i\ell\rangle,
\qquad
Z_i=\sum_\ell e^{-\epsilon_{i\ell}/T}.
\label{method:thermal_average}
\end{equation}
The local moments are updated iteratively until self-consistency is reached.
The free energy is evaluated as
\begin{equation}
F=-T\sum_i\ln Z_i+E_{\mathrm{dc}}.
\label{method:free_energy}
\end{equation}

All calculations are performed on a $6\times6$ square lattice with periodic boundary conditions at $T=0.05$, where $J_1=1$ sets the energy scale.
The convergence tolerance is $10^{-6}$.
To reduce the dependence on initial conditions, we use at least 793 initial configurations, including single-$Q$, double-$Q$, conical, SkL, ferromagnetic, and random states, as well as solutions obtained at nearby
parameter points\footnote{Further details on the initial configurations
can be found in Ref.~\cite{zha2026effects}.}.
For each parameter set, the converged state with the lowest free energy is adopted.
Magnetic phases are identified from the real-space moment configurations, the component-resolved structure factors, scalar chirality, and $N_{\mathrm{sk}}$, as shown below.

\subsection{Magnetic structure factors}

We characterize the magnetic modulations using the Fourier components of the self-consistent mean moments,
\begin{equation}
\mathbf m_{\mathbf q}
=
\frac{1}{\sqrt N}
\sum_i
\mathbf m_i e^{-i\mathbf q\cdot\mathbf r_i},
\label{method:moment_fourier}
\end{equation}
where $N$ is the number of lattice sites in the computational cell.
The component-resolved magnetic structure factor is defined by
\begin{equation}
J^\mu(\mathbf q)=|m_{\mathbf q}^{\mu}|^2,
\qquad \mu=x,y,z,
\label{method:structure_factor}
\end{equation}
and we use
\begin{equation}
\begin{aligned}
J^{xy}(\mathbf q)&=J^x(\mathbf q)+J^y(\mathbf q),\\
J(\mathbf q)&=J^{xy}(\mathbf q)+J^z(\mathbf q).
\end{aligned}
\label{method:structure_components}
\end{equation}
These quantities are used to identify the dominant wave vectors and to
distinguish different single-$Q$, multiple-$Q$, square-SkL, and
rectangular-SkL states.
The $\mathbf q=\mathbf 0$ component represents the uniform moment,
with $\mathbf m_{\mathbf 0}=\sqrt N\,\langle\mathbf J\rangle$ and
$J^\mu(\mathbf 0)=N\langle J^\mu\rangle^2$.

\subsection{Scalar chirality and topological charge}

To characterize noncoplanar magnetic order, we calculate the local scalar
chirality~\cite{Wen_PhysRevB.39.11413, Kawamura_PhysRevLett.68.3785},
\begin{equation}
\chi_i
=
\frac12
\sum_{s,s'=\pm1}
ss'\,
\mathbf m_i\cdot
\left(
\mathbf m_{i+s\hat{\mathbf x}}
\times
\mathbf m_{i+s'\hat{\mathbf y}}
\right),
\label{method:chirality_local}
\end{equation}
and its spatial average
\begin{equation}
\langle\chi\rangle
=
\frac1N\sum_i\chi_i.
\label{method:chirality_average}
\end{equation}
The spatial distribution of $\chi_i$ reflects local noncoplanarity, whereas $\langle\chi\rangle$ quantifies the net scalar chirality.

The topological charge is evaluated from the normalized local moments
$\hat{\mathbf m}_i=\mathbf m_i/|\mathbf m_i|$ using the discrete
geometrical prescription of Ref.~\cite{BERG1981412}.
For each site, we define
\begin{equation}
\Omega_i
=
\sum_{s,s'=\pm1}
\arctan\left[
\frac{
2ss'\,\hat{\mathbf m}_i\cdot
(\hat{\mathbf m}_j\times\hat{\mathbf m}_k)
}{
|\hat{\mathbf m}_i+\hat{\mathbf m}_j+\hat{\mathbf m}_k|^2-1
}
\right],
\label{method:solid_angle}
\end{equation}
where $j=i+s\hat{\mathbf x}$ and $k=i+s'\hat{\mathbf y}$.
We evaluate the total topological charge over the $6\times6$ computational cell and the skyrmion number per primitive magnetic unit cell as
\begin{equation}
\begin{aligned}
N_{\mathrm{sk}}^{\mathrm{tot}}
&=\frac{1}{4\pi}\sum_{i=1}^{36}\Omega_i,\\
N_{\mathrm{sk}}
&=\frac{N_{\mathrm{sk}}^{\mathrm{tot}}}{n_{\mathrm{mag}}}.
\end{aligned}
\label{method:Nsk}
\end{equation}
Here, $n_{\mathrm{mag}}$ is the number of primitive magnetic unit cells contained in the computational cell.
The primitive magnetic unit cell is determined from translations that reproduce the full local-moment configuration, including both its direction and magnitude.
For $n_{\mathrm{mag}}=2$, the normalization factor is $1/2$. 
The primitive translations and $n_{\mathrm{mag}}$ for the representative SkLs are given in Sec.~\ref{sec:results}.

\section{Results}
\label{sec:results}
\subsection{Magnetic phases and skyrmion lattices in the single-orbital $J_1$--DMI model}
\label{results:single_orbital}
\begin{figure}[t]
\centering
\includegraphics[width=\columnwidth]{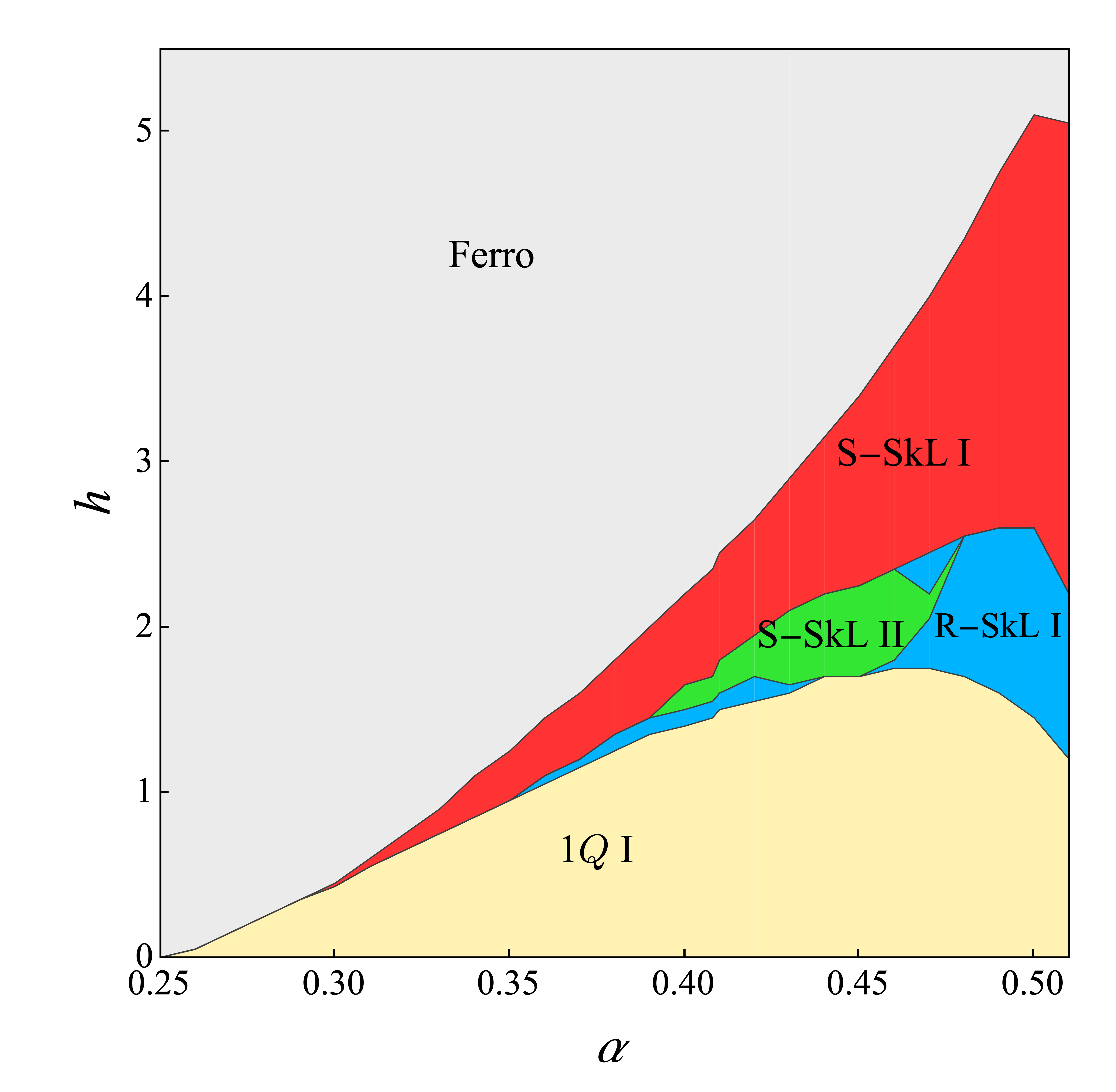}
\caption{
Low-temperature mean-field phase diagram of the single-orbital $J_1$--DMI model in the $\alpha$--$h$ plane 
for $J_1=1$, $J_2=0$, $|D|=\sqrt{6}$, and $T=0.05$. 
The phase diagram is sampled in steps of $0.01$ in $\alpha$ and $0.05$ in $h$, with an additional scan at $\alpha=0.408$.
At $\alpha=0.30$, the narrow S-SkL~I region is resolved using a finer field step of $0.01$.
The magnetic phases are the single-$Q$ vertical state ($1Q$~I), S-SkL~I, S-SkL~II, R-SkL~I, and the field-polarized state (Ferro).
}
\label{fig:single_phase}
\end{figure}

We first examine the single-orbital model to clarify how the crystal-field wave-function composition controls DMI-induced magnetic textures.
We set $J_1=1$, $J_2=0$, $|D|=\sqrt6$, and $T=0.05$, and construct the mean-field phase diagram in the $\alpha$–$h$ plane.
Figure~\ref{fig:single_phase} shows five phases: a single-$Q$ vertical state ($1Q$~I), two square skyrmion lattices (S-SkL~I and S-SkL~II), a rectangular skyrmion lattice (R-SkL~I), and the field-polarized state.

Although S-SkL~I and S-SkL~II share square-lattice symmetry, they are distinguished by their magnetic modulation periods, whereas R-SkL~I exhibits a rectangular arrangement of skyrmions.
Thus, these SkL phases differ in their magnetic periodicities and/or spatial symmetries.

The stability of the SkLs changes markedly with $\alpha$.
At small $\alpha$, the magnetic field drives the single-$Q$ state directly into the field-polarized state, whereas S-SkL~I emerges near $\alpha=0.30$ and its stability region expands substantially with
increasing $\alpha$.
By contrast, S-SkL~II and R-SkL~I show a more intricate dependence on $\alpha$.
S-SkL~II is stabilized over an intermediate range of $\alpha$, while R-SkL~I appears in neighboring regions and competes with S-SkL~II.

This competition is evident in representative field scans. 
At $\alpha=0.408$, increasing the field yields the sequence single-$Q$ $\rightarrow$ R-SkL~I $\rightarrow$ S-SkL~II $\rightarrow$ S-SkL~I $\rightarrow$ field-polarized.
At $\alpha=0.45$, R-SkL~I is absent and the single-$Q$ state transforms directly into S-SkL~II, whereas at $\alpha=0.50$, S-SkL~II is replaced by R-SkL~I.
A particularly notable case occurs near $\alpha=0.47$, where R-SkL~I appears in two separate field intervals with S-SkL~II in between, resulting in reentrant behavior.
Thus, varying the crystal-field wave-function composition changes both the stability of the SkLs and the competition among states with different magnetic periodicities and spatial symmetries.

The pronounced $\alpha$ dependence originates from the crystal-field matrix elements in Eq.~\eqref{model:projection}.
Within the parameter range considered here, increasing $\alpha$ changes the projected doublet from easy-axis to easy-plane character, with the in-plane and out-of-plane matrix elements becoming equal in magnitude near $\alpha\simeq0.408$.
Thus, $\alpha$ continuously tunes the effective magnetic anisotropy.
At the same time, these matrix elements enter the exchange, DMI, and
Zeeman terms, so that changing $\alpha$ also modifies their effective coupling strengths.
The phase evolution therefore reflects a correlated change of these energy scales rather than the variation of an independent anisotropy parameter.
This differs from spin-only models in which magnetic anisotropy is introduced as an independent phenomenological coupling.
Here, both the anisotropy and the effective coupling strengths originate from the crystal-field wave function.

\begin{figure*}[t]
\centering
\includegraphics[width=\textwidth]{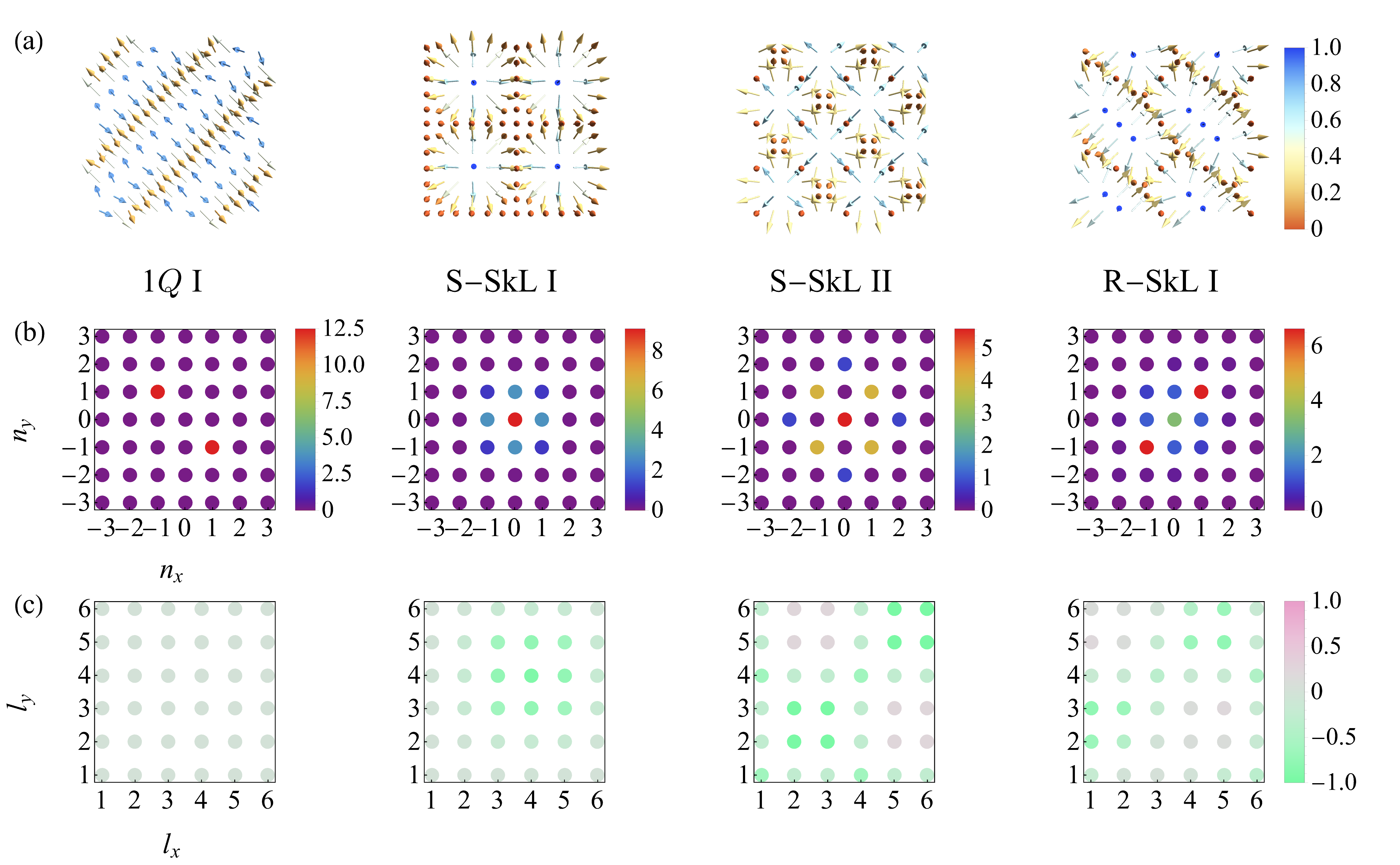}
\caption{Representative magnetic configurations in the
single-orbital $J_1$--DMI model.
From left to right, the columns
correspond to the single-$Q$ vertical state
($1Q$~I), S-SkL~I, S-SkL~II, and R-SkL~I,
while the rows show
(a) the real-space mean moments,
with color representing the normalized polar angle $\theta/\pi$,
(b) the total magnetic structure factor $J(\mathbf q)$,
and (c) the local scalar chirality $\chi_i$.
Panel (a) displays a $2\times2$ periodic tiling of the $6\times6$ computational cell, while panel (c) shows one cell.
The upper and right boundary points in panel (b) are periodic copies of those on the opposite boundaries.
The parameters are $J_1=1$, $J_2=0$,
$|D|=\sqrt{6}$, and $T=0.05$.
The total topological charges evaluated over the
$6\times6$ computational cell have magnitudes
$|N_{\mathrm{sk}}^{\mathrm{tot}}|=1$ for S-SkL~I
and R-SkL~I, and
$|N_{\mathrm{sk}}^{\mathrm{tot}}|=2$ for S-SkL~II.
The primitive magnetic unit cells and the corresponding
skyrmion numbers per magnetic unit cell are discussed
in the text.}
\label{fig:single_structures}
\end{figure*}

Figure~\ref{fig:single_structures} compares representative real-space textures, structure factors, and local scalar chiralities.
The single-$Q$ state is dominated by one pair of opposite diagonal wave vectors, whereas the SkLs involve several Fourier components and form noncoplanar multiple-$Q$ textures.

S-SkL~I and S-SkL~II both preserve the fourfold symmetry of the square lattice but are characterized by distinct
Fourier contents and magnetic periods.
S-SkL~I has a sizable uniform component, whereas S-SkL~II exhibits stronger contributions at the combination wave vectors $\mathbf Q_3=(2Q,0)$ and $\mathbf Q_4=(0,2Q)$, resulting in a shorter magnetic period.

To clarify this difference, we determine the primitive magnetic unit cells of representative configurations at $\alpha=0.408$.
For S-SkL~I at $h=2.00$, primitive translations can be chosen as $(6,0)$ and $(0,6)$, in units of the lattice constant, giving a magnetic unit cell containing 36 sites.
For S-SkL~II at $h=1.65$, the primitive translations $(3,3)$ and $(-3,3)$ instead enclose 18 sites.
The $6\times6$ computational cell therefore contains one primitive magnetic cell of S-SkL~I and two of S-SkL~II.
Their total topological-charge magnitudes are $|N_{\mathrm{sk}}^{\mathrm{tot}}|=1$ and 2, respectively, whereas both have $|N_{\mathrm{sk}}|=1$ per primitive magnetic unit cell.
Thus, the difference in total charge reflects their different magnetic periodicities.
For the representative R-SkL~I configuration at $(\alpha,h)=(0.408,1.50)$, primitive translations can
also be chosen as $(6,0)$ and $(0,6)$.
Its primitive magnetic unit cell contains 36 sites and has $|N_{\mathrm{sk}}|=1$.

Despite having the same skyrmion number per primitive magnetic unit cell, S-SkL~I and R-SkL~I exhibit qualitatively different spatial symmetries. 
In contrast to the fourfold-symmetric S-SkL~I, R-SkL~I is characterized by unequal diagonal structure-factor intensities, $J(\mathbf Q_1)\neq J(\mathbf Q_2)$, despite the fourfold rotational symmetry of the Hamiltonian. 
This rectangular SkL therefore spontaneously breaks the in-plane $C_4$ symmetry, demonstrating that the skyrmion number alone cannot distinguish these states. 
Experimentally, a rhombic SkL with broken fourfold rotational symmetry has been observed in the centrosymmetric tetragonal magnet EuAl$_4$~\cite{Shang_PhysRevB.103.L020405,kaneko2021charge,takagi2022square,Gen_PhysRevB.107.L020410,Arai2026,Gen2026}, while multiple skyrmion phases with distinct symmetries have been reported in the centrosymmetric magnet GdRu$_2$Ge$_2$~\cite{yoshimochi2024multi,Rathnaweera2026}. 
The present results demonstrate that symmetry-lowered SkLs can also emerge in noncentrosymmetric systems with DMI.

Previous studies of square SkLs have emphasized anisotropic interactions or higher-harmonic couplings as mechanisms for selecting fourfold or rectangular multiple-$Q$ textures~\cite{Yi_PhysRevB.80.054416, Hayami_PhysRevB.105.174437, hayami2023widely}.
The present result shows that, even for the nearest-neighbor exchange--DMI model considered here, the crystal-field wave function itself can tune the competition between square and spontaneously fourfold-symmetry-broken SkLs.

\subsection{Effects of second-neighbor exchange}
\label{results:second_neighbor}
\begin{figure*}[!tp]
\centering
\includegraphics[width=0.95\textwidth]
{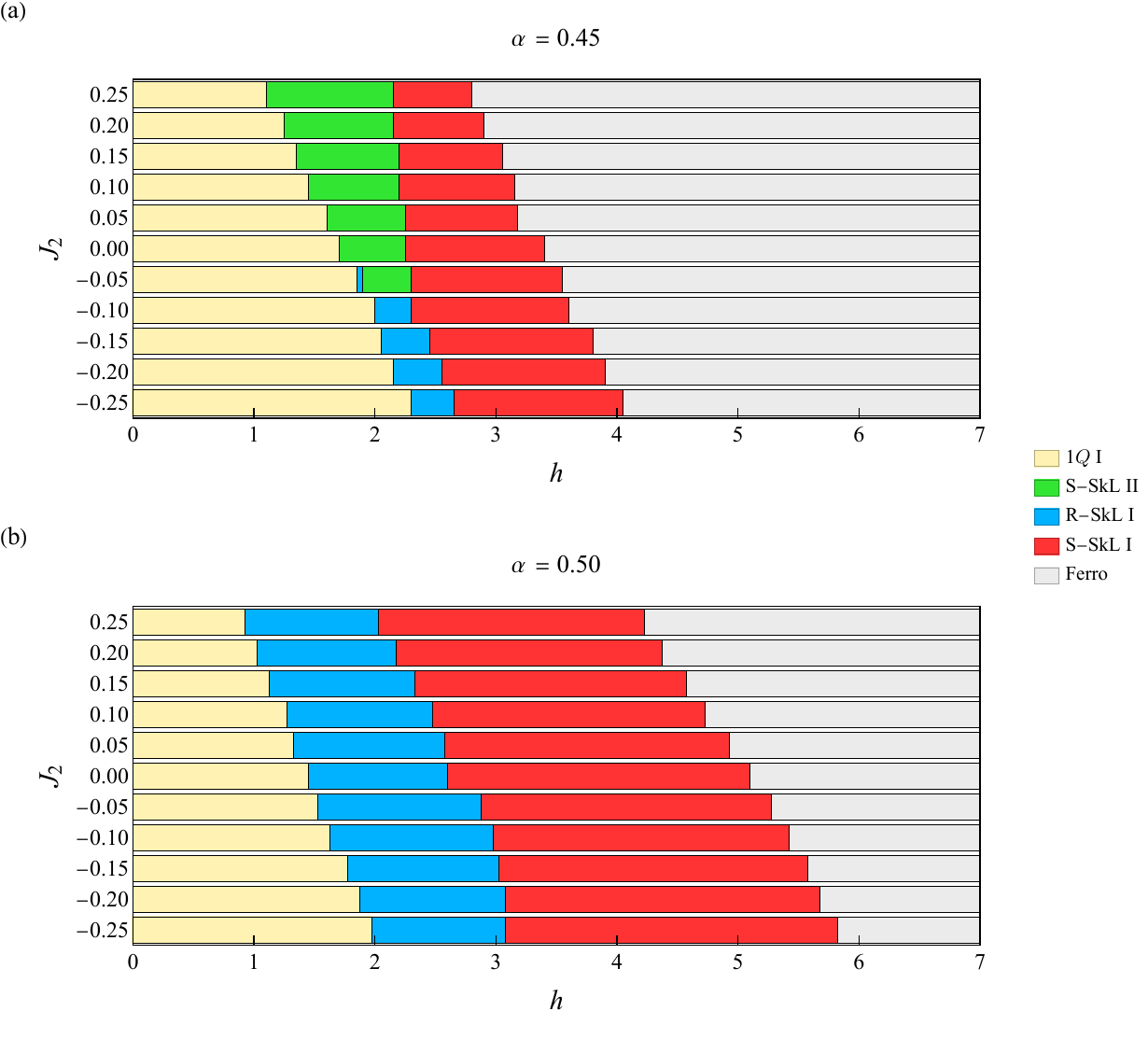}

\caption{
Low-temperature mean-field phase diagrams of the single-orbital
$J_1$--$J_2$--DMI model in the $J_2$--$h$ plane at
(a) $\alpha=0.45$ and (b) $\alpha=0.50$.
We set $J_1=1$ and $T=0.05$, while the DMI strength is varied
according to Eq.~\eqref{model:Dpath}.
The phase diagrams are sampled over
$-0.25\leq J_2\leq0.25$ in steps of $0.05$.
Each row shows a field scan at a fixed $J_2$,
with a field step of $0.05$.
The $J_2=0$ rows correspond to the nearest-neighbor model.
The phases are the single-$Q$ vertical state
($1Q$~I), S-SkL~I, S-SkL~II, R-SkL~I,
and the field-polarized state (Ferro).
The narrow blue segment at $J_2=-0.05$ in panel (a)
denotes R-SkL~I.
}
\label{fig:second_neighbor_phase}
\end{figure*}

We next examine how the SkLs found in the nearest-neighbor model respond to an additional exchange channel.
We introduce $J_2$ and vary $D$ according to Eq.~\eqref{model:Dpath}, so that the diagonal reference wave vectors remain stationary.
Figure~\ref{fig:second_neighbor_phase} shows the $J_2$--$h$ phase diagrams for $\alpha=0.45$ and $0.50$.
These two values are chosen because the $J_2=0$ model favors S-SkL~II and R-SkL~I, respectively, at intermediate fields.

At $\alpha=0.45$, positive $J_2$ expands the S-SkL~II region mainly toward lower fields.
Negative $J_2$ progressively suppresses S-SkL~II.
At $J_2=-0.05$, a narrow R-SkL~I region appears between the $1Q$~I state and S-SkL~II.
For the sampled values $J_2\leq-0.10$, S-SkL~II is absent, and the phase sequence with increasing field is $1Q$~I, R-SkL~I, S-SkL~I, and Ferro.
Thus, the additional exchange channel can switch the preferred intermediate-field texture from S-SkL~II, characterized by a shorter magnetic period, to the spontaneously fourfold-symmetry-broken R-SkL~I.

At $\alpha=0.50$, by contrast, R-SkL~I remains stable throughout the sampled $J_2$ range.
Changing $J_2$ mainly shifts the field positions of R-SkL~I and S-SkL~I rather than changing their sequence.
The phase sequence remains $1Q$~I, R-SkL~I, S-SkL~I, and Ferro.
Compared with $J_2=0$, positive $J_2$ shifts the phase boundaries to lower fields, while negative $J_2$ shifts them to higher fields.
The different responses at $\alpha=0.45$ and $0.50$ demonstrate that the effect of an additional exchange interaction depends strongly on the underlying crystal-field wave function.

Since the DMI strength is varied together with $J_2$ according to Eq.~\eqref{model:Dpath}, the phase evolution does not represent the effect of $J_2$ alone.
This parameter path is chosen to keep the diagonal reference wave vectors $\mathbf Q_1$ and $\mathbf Q_2$ as stationary points of the interaction kernel while varying the balance of the exchange and DMI interactions.

\subsection{Two-orbital model: Crossover to the single-orbital limit}
\label{results:single_orbital_limit}

\begin{figure*}[t]
\centering
\includegraphics[width=\textwidth]
{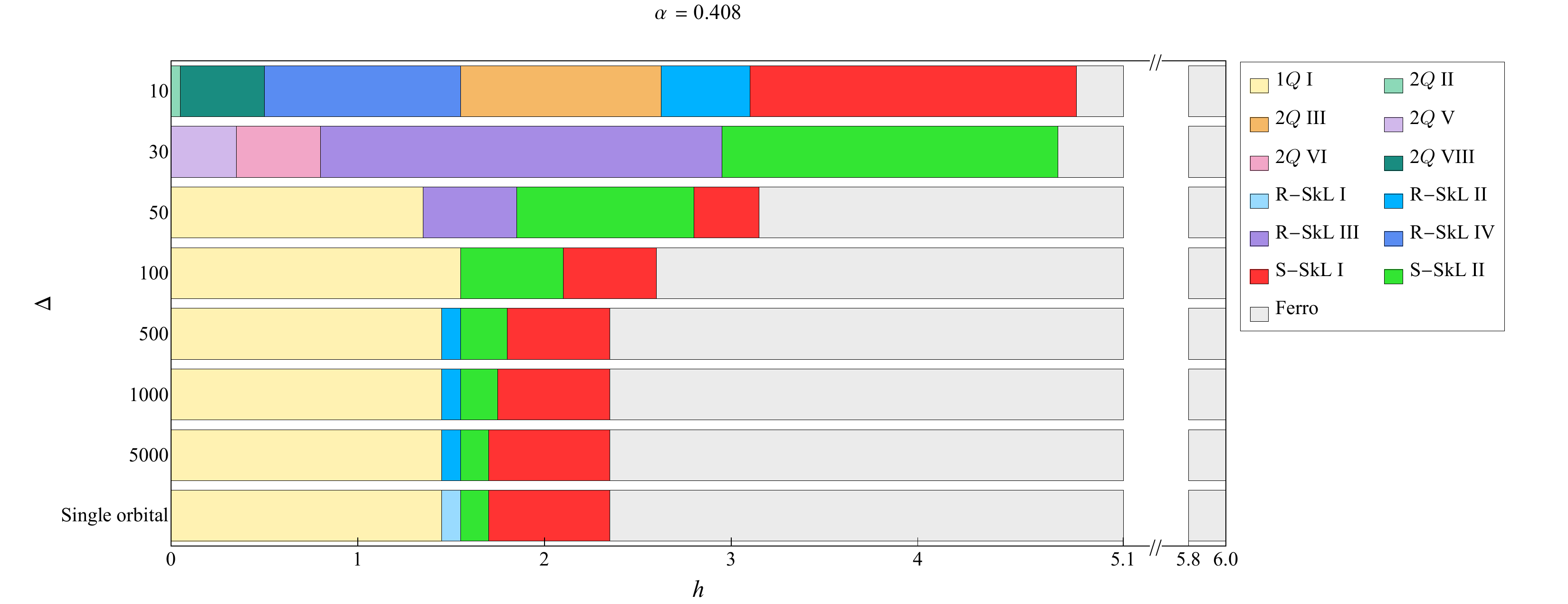}
\caption{
Comparison of low-temperature mean-field field scans for the two-orbital model at $\alpha=0.408$ and several crystal-field splittings $\Delta=10$, 30, 50, 100, 500, 1000, and 5000, together with the single-orbital reference at the same $\alpha$.
Each row corresponds to a separate field scan.
The magnetic field is sampled in steps of $0.05$.
The parameters are $J_1=1$, $J_2=0$, $|D|=\sqrt{6}$, and $T=0.05$.
The colors indicate the magnetic phases shown in the legend, with Ferro denoting the field-polarized state.
For $\Delta=500$, 1000, and 5000, the rectangular SkL is R-SkL~II, whereas the single-orbital model exhibits R-SkL~I.
The horizontal axis is broken between $h=5.1$ and $5.8$.
}
\label{fig:orbital_crossover}
\end{figure*}

We next restore the second $\Gamma_7$ doublet and examine how the single-orbital description emerges as the crystal-field splitting $\Delta$ increases.
We fix $\alpha=0.408$, $J_1=1$, $J_2=0$, $|D|=\sqrt6$, and $T=0.05$. 
For large positive $\Delta$, the second doublet is energetically suppressed, whereas decreasing $\Delta$ enhances its admixture through the off-diagonal matrix elements of $\mathbf J$.

Figure~\ref{fig:orbital_crossover} compares representative field scans from $\Delta=10$ to $5000$ with the single-orbital result.
The phase boundaries approach those of the single-orbital model as $\Delta$ increases, confirming the expected crossover.
However, the convergence of the phase boundaries is faster than that of the detailed magnetic textures.
For $\Delta=500$, 1000, and 5000, the intermediate rectangular phase is classified as R-SkL~II, whereas the strict single-orbital model contains R-SkL~I.
The two phases differ in the Fourier components responsible for the broken $C_4$ symmetry: R-SkL~I has unequal intensities at $\mathbf Q_1$ and $\mathbf Q_2$, whereas in R-SkL~II these intensities remain equal and the inequivalence instead appears at the subsidiary wave vectors.
An additional calculation at $\Delta=10000$, not shown in Fig.~\ref{fig:orbital_crossover}, recovers the R-SkL~I texture of the single-orbital model within the present numerical resolution.

This distinction shows that the crossover to the single-orbital limit does not occur uniformly for all properties of the ordered state. 
As $\Delta$ increases, the excited crystal-field doublet becomes progressively less involved, so that the overall phase sequence and phase boundaries approach those of the single-orbital model.
Nevertheless, even a small admixture of the excited doublet can modify the relative weights of different Fourier components through the interorbital matrix elements of the angular momentum.
Consequently, the gross phase structure can become nearly indistinguishable from that of the single-orbital model while the detailed magnetic texture, such as the distribution of the Fourier intensities responsible for the broken $C_4$ symmetry, remains different.
The validity of a single-orbital description should therefore be assessed not only from the phase boundaries but also from the detailed Fourier content and real-space magnetic texture.

As $\Delta$ is reduced further, the phase sequence changes qualitatively.
Additional rectangular SkLs and non-skyrmion double-$Q$ states appear, and the ordering wave vectors can shift from diagonal to axial directions.
At $\Delta=10$, increasing the field produces the sequence $2Q$~II, $2Q$~VIII, R-SkL~IV, $2Q$~III, R-SkL~II, S-SkL~I, and Ferro.
Since $J_1$, $J_2$, and $D$ are fixed in these scans, these changes cannot be attributed to a modification of the bare exchange--DMI dispersion.
They instead originate from the enlarged local Hilbert space and the interorbital matrix elements of the angular momentum.

\subsection{Magnetic phases at small crystal-field splitting}
\label{results:small_splitting}

\begin{figure}[t]
\centering
\includegraphics[width=\columnwidth]{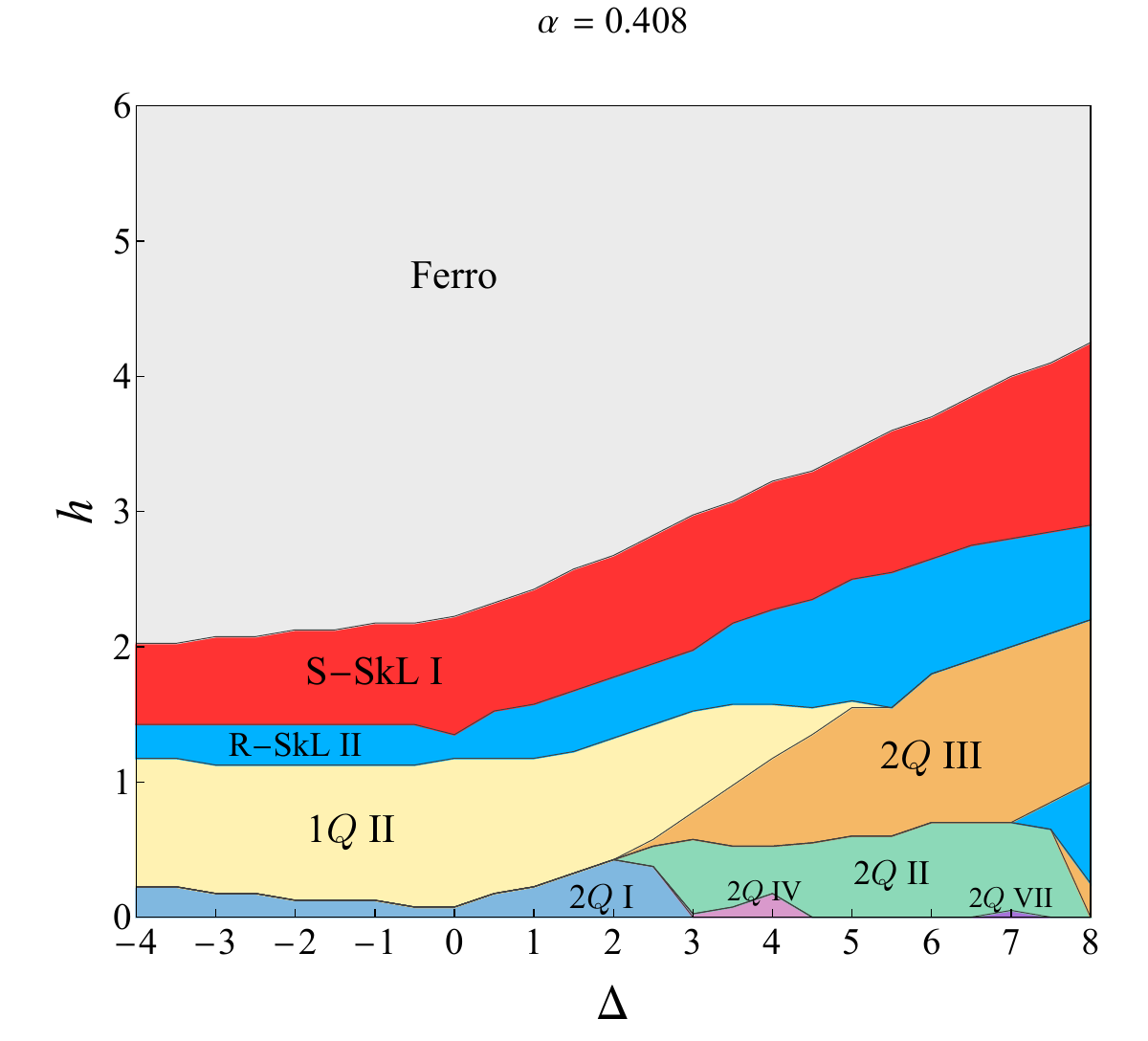}
\caption{
Low-temperature mean-field phase diagram of the two-orbital model in the $\Delta$--$h$ plane at $\alpha=0.408$, $J_1=1$, $J_2=0$, $|D|=\sqrt{6}$, and $T=0.05$. 
The displayed ranges are $-4\leq\Delta\leq8$ and $0\leq h\leq6$.
The sampling intervals are $0.5$ in $\Delta$ and $0.05$ in $h$.
Ferro denotes the field-polarized state, and the other phase labels follow the classification in Table~\ref{tab:magnetic_phase_classification}.
}
\label{fig:small_splitting_phase}
\end{figure}

We finally focus on the regime where both crystal-field doublets participate substantially in the magnetic state.
Figure~\ref{fig:small_splitting_phase} shows the $\Delta$--$h$ phase diagram for $\alpha=0.408$.
Compared with the single-orbital phase diagram, the small-$\Delta$ regime contains a much wider variety of magnetic states, including an axial single-$Q$ state ($1Q$~II), several double-$Q$ states ($2Q$~I, $2Q$~II, $2Q$~III, $2Q$~IV, and $2Q$~VII), and the SkLs R-SkL~II and S-SkL~I.

For negative and weakly positive $\Delta$, the low-field region is dominated by the axial $1Q$~II state.
At intermediate fields, R-SkL~II and S-SkL~I occupy broad regions. 
On the positive-$\Delta$ side, several double-$Q$ states intervene between the low-field and SkL phases.
Near $\Delta=8$, R-SkL~II becomes reentrant, with the $2Q$~III state separating its two field intervals.

For the following discussion, we use the wave vectors $\mathbf Q_1$--$\mathbf Q_6$ defined in Sec.~\ref{model:parameter_choice}, where $\mathbf Q_3$ and $\mathbf Q_4$ are combination wave vectors of $\mathbf Q_1$ and $\mathbf Q_2$, while $\mathbf Q_5$ and $\mathbf Q_6$ are axial modes compatible with the $6\times6$ computational cell.
Further details of the momentum-space interactions and the parameter path for varying $J_2$ are given in Appendices~\ref{app:fourier_interaction} and~\ref{app:J2_parameter_path}, respectively.

\begin{table*}[!tp]
\caption{
Classification of the magnetic phases according to the structure-factor components at $\mathbf Q_1$--$\mathbf Q_6$.
Here, $J_{\mathbf Q_\eta}^{xy}=J_{\mathbf Q_\eta}^{x}+J_{\mathbf Q_\eta}^{y}$.
Within each wave-vector pair, the upper and lower lines indicate the relations for the $xy$ and $z$ components, respectively; these relations do not imply equal intensities between the two channels.
Equalities and zeros are defined within numerical accuracy, and $>0$ denotes a nonzero intensity.
The notation $A>B$ allows $B=0$, whereas $A>B>0$ indicates that both components are nonzero.
The last column lists the magnitude $|N_{\mathrm{sk}}^{\mathrm{tot}}|$ of the total topological charge evaluated over the $6\times6$ computational cell.
The table summarizes phases identified in both the single-orbital and two-orbital calculations, including those outside the parameter window of Fig.~\ref{fig:small_splitting_phase}.
Symmetry-related domains are regarded as equivalent under the corresponding transformation of the wave-vector labels.
}
\label{tab:magnetic_phase_classification}
\centering
\begingroup
\small
\newcommand{\phxy}[1]{J_{\mathbf{Q}_{#1}}^{xy}}
\newcommand{\phz}[1]{J_{\mathbf{Q}_{#1}}^{z}}
\newcommand{\phpair}[2]{%
\begin{tabular}{@{}c@{}}#1\\#2\end{tabular}%
}
\setlength{\tabcolsep}{4pt}
\renewcommand{\arraystretch}{1.08}
\resizebox{0.85\textwidth}{!}{%
\begin{tabular}{lcccc}
\hline\hline
phase
& $J_{\mathbf Q_1},J_{\mathbf Q_2}\;([110])$
& $J_{\mathbf Q_3},J_{\mathbf Q_4}\;([100])$
& $J_{\mathbf Q_5},J_{\mathbf Q_6}\;([100])$
& $|N_{\mathrm{sk}}^{\mathrm{tot}}|$ \\
\hline
$1Q$~I
& \phpair{$\phxy{1}>0,\quad \phxy{2}=0$}
          {$\phz{1}>0,\quad \phz{2}=0$}
& \phpair{$\phxy{3}=\phxy{4}=0$}
          {$\phz{3}=\phz{4}=0$}
& \phpair{$\phxy{5}=\phxy{6}=0$}
          {$\phz{5}=\phz{6}=0$}
& $0$ \\
$1Q$~II
& \phpair{$\phxy{1}=\phxy{2}=0$}
          {$\phz{1}=\phz{2}=0$}
& \phpair{$\phxy{3}>0,\quad \phxy{4}=0$}
          {$\phz{3}>0,\quad \phz{4}=0$}
& \phpair{$\phxy{5}>0,\quad \phxy{6}=0$}
          {$\phz{5}>0,\quad \phz{6}=0$}
& $0$ \\
\hline
S-SkL~I
& \phpair{$\phxy{1}=\phxy{2}>0$}
          {$\phz{1}=\phz{2}>0$}
& \phpair{$\phxy{3}=\phxy{4}>0$}
          {$\phz{3}=\phz{4}>0$}
& \phpair{$\phxy{5}=\phxy{6}>0$}
          {$\phz{5}=\phz{6}>0$}
& $1$ \\
S-SkL~II
& \phpair{$\phxy{1}=\phxy{2}>0$}
          {$\phz{1}=\phz{2}>0$}
& \phpair{$\phxy{3}=\phxy{4}>0$}
          {$\phz{3}=\phz{4}>0$}
& \phpair{$\phxy{5}=\phxy{6}=0$}
          {$\phz{5}=\phz{6}=0$}
& $2$ \\
R-SkL~I
& \phpair{$\phxy{1}>\phxy{2}$}
          {$\phz{1}>\phz{2}$}
& \phpair{$\phxy{3}=\phxy{4}>0$}
          {$\phz{3}=\phz{4}>0$}
& \phpair{$\phxy{5}=\phxy{6}>0$}
          {$\phz{5}=\phz{6}>0$}
& $1$ \\
R-SkL~II
& \phpair{$\phxy{1}=\phxy{2}>0$}
          {$\phz{1}=\phz{2}>0$}
& \phpair{$\phxy{3}>\phxy{4}$}
          {$\phz{3}>\phz{4}$}
& \phpair{$\phxy{5}>\phxy{6}$}
          {$\phz{5}>\phz{6}$}
& $1$ \\
R-SkL~III
& \phpair{$\phxy{1}=\phxy{2}>0$}
          {$\phz{1}=\phz{2}>0$}
& \phpair{$\phxy{3}>\phxy{4}$}
          {$\phz{3}>\phz{4}$}
& \phpair{$\phxy{5}=\phxy{6}=0$}
          {$\phz{5}=\phz{6}=0$}
& $2$ \\
R-SkL~IV
& \phpair{$\phxy{1}=\phxy{2}>0$}
          {$\phz{1}=\phz{2}>0$}
& \phpair{$\phxy{3}>\phxy{4}$}
          {$\phz{3}>\phz{4}$}
& \phpair{$\phxy{5}>\phxy{6}$}
          {$\phz{5}>\phz{6}$}
& $2$ \\
\hline
$2Q$~I
& \phpair{$\phxy{1}>\phxy{2}>0$}
          {$\phz{1}>\phz{2}>0$}
& \phpair{$\phxy{3}=\phxy{4}>0$}
          {$\phz{3}=\phz{4}>0$}
& \phpair{$\phxy{5}=\phxy{6}=0$}
          {$\phz{5}=\phz{6}=0$}
& $0$ \\
$2Q$~II
& \phpair{$\phxy{1}>\phxy{2}>0$}
          {$\phz{1}>\phz{2}>0$}
& \phpair{$\phxy{3}>\phxy{4}>0$}
          {$\phz{3}>\phz{4}>0$}
& \phpair{$\phxy{5}>\phxy{6}>0$}
          {$\phz{5}>\phz{6}>0$}
& $0$ \\
$2Q$~III
& \phpair{$\phxy{1}>\phxy{2}$}
          {$\phz{1}>\phz{2}$}
& \phpair{$\phxy{3}=\phxy{4}=0$}
          {$\phz{3}=\phz{4}=0$}
& \phpair{$\phxy{5}=\phxy{6}=0$}
          {$\phz{5}=\phz{6}=0$}
& $0^{\mathrm a}$ \\
$2Q$~IV
& \phpair{$\phxy{1}>\phxy{2}>0$}
          {$\phz{1}>\phz{2}>0$}
& \phpair{$\phxy{3}>\phxy{4}>0$}
          {$\phz{3}>\phz{4}>0$}
& \phpair{$\phxy{5}\simeq\phxy{6}>0$}
          {$\phz{5}\simeq\phz{6}>0$}
& $0$ \\
$2Q$~V
& \phpair{$\phxy{1}=\phxy{2}>0$}
          {$\phz{1}=\phz{2}>0$}
& \phpair{$\phxy{3}=\phxy{4}=0$}
          {$\phz{3}=\phz{4}=0$}
& \phpair{$\phxy{5}>\phxy{6}$}
          {$\phz{5}>\phz{6}$}
& $0$ \\
$2Q$~VI
& \phpair{$\phxy{1}=\phxy{2}>0$}
          {$\phz{1}=\phz{2}>0$}
& \phpair{$\phxy{3}>\phxy{4}$}
          {$\phz{3}>\phz{4}$}
& \phpair{$\phxy{5}=\phxy{6}=0$}
          {$\phz{5}=\phz{6}=0$}
& $0$ \\
$2Q$~VII
& \phpair{$\phxy{1}>\phxy{2}$}
          {$\phz{1}>\phz{2}$}
& \phpair{$\phxy{3}=\phxy{4}=0$}
          {$\phz{3}=\phz{4}=0$}
& \phpair{$\phxy{5}>0,\quad \phxy{6}=0$}
          {$\phz{5}>0,\quad \phz{6}=0$}
& $0$ \\
\hline\hline
\end{tabular}%
}

\par\smallskip
\begin{minipage}{0.85\textwidth}
\footnotesize
\raggedright
\textsuperscript{a}For the $2Q$~III state, the geometrical
topological charge evaluated over the $6\times6$ computational
cell depends on the triangulation of the square plaquettes,
yielding $N_{\mathrm{sk}}^{\mathrm{tot}}=0$ or $-3$.
Zero is listed here as a reporting convention, with this
triangulation dependence explicitly noted.
\par
\end{minipage}

\endgroup
\end{table*}

\subsection{Magnetic structures in the two-orbital model}

\subsubsection{Square and rectangular skyrmion lattices}

\begin{figure*}[t]
\centering
\includegraphics[width=\textwidth]
{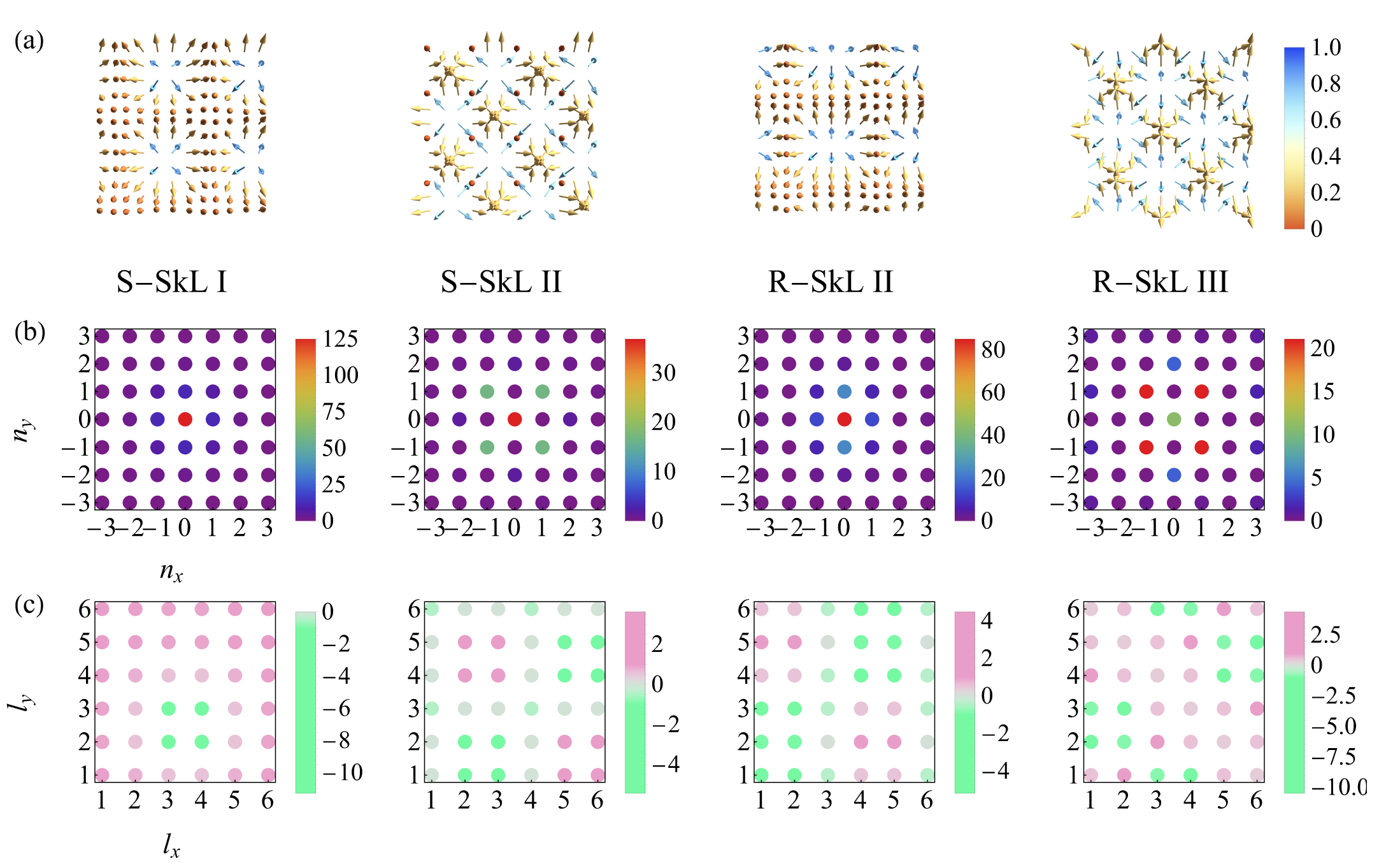}
\caption{
Representative SkLs obtained in the two-orbital model.
From left to right, the columns correspond to S-SkL~I, S-SkL~II, R-SkL~II, R-SkL~III, and R-SkL~IV.
The rows show (a) the real-space mean moments, with color representing the normalized polar angle $\theta/\pi$, (b) the total magnetic structure factor $J(\mathbf q)$ defined in Eq.~\eqref{method:structure_components}, where $\mathbf q=Q(n_x,n_y)$ and $Q=\pi/3$, and (c) the local scalar chirality $\chi_i$ defined in Eq.~\eqref{method:chirality_local}.
Panel (a) displays a $2\times2$ periodic tiling of the $6\times6$ computational cell, while panel (c) shows one cell.
The upper and right boundary points in panel (b) are periodic copies of those on the opposite boundaries.
The structure-factor and chirality panels use separate color scales.
From left to right, the magnitudes of the total topological charges evaluated over the $6\times6$ computational cell are $|N_{\mathrm{sk}}^{\mathrm{tot}}|=1$, 2, 1, 2, and 2, respectively.
}
\label{fig:two_orbital_skyrmion}
\end{figure*}

Figure~\ref{fig:two_orbital_skyrmion} compares representative SkLs obtained in the two-orbital model.
The two square SkLs preserve the equivalence between symmetry-related wave vectors but differ in their subsidiary Fourier components and magnetic periodicities.
In particular, S-SkL~II has a shorter magnetic period than S-SkL~I, consistent with its enhanced Fourier components at the combination wave vectors.
When evaluated over the $6\times6$ computational cell, S-SkL~I and S-SkL~II have total topological-charge magnitudes $|N_{\mathrm{sk}}^{\mathrm{tot}}|=1$ and 2, respectively.
Thus, the distinction between the two square SkLs primarily reflects a reorganization of the Fourier content and real-space magnetic texture, including a change in magnetic periodicity.

The rectangular SkLs instead break the fourfold equivalence in different parts of momentum space.
For R-SkL~I, the asymmetry already appears between the diagonal components at $\mathbf Q_1$ and $\mathbf Q_2$.
For R-SkL~II, these diagonal components remain equivalent, while the asymmetry appears at the combination and axial wave vectors.
Thus, R-SkL~I and R-SkL~II have the same total topological-charge magnitude $|N_{\mathrm{sk}}^{\mathrm{tot}}|=1$ but represent distinct symmetry-breaking patterns.
This demonstrates that the topological charge alone is insufficient to classify the rectangular SkLs; the momentum-space distribution of the broken $C_4$ symmetry provides an additional distinction.
The difference is also manifested in their real-space textures, despite their identical net topological charge.

To determine the primitive magnetic unit cells, we examine representative configurations at $\alpha=0.408$.
For S-SkL~I at $(\Delta,h)=(10,3.50)$, R-SkL~II at $(10,2.85)$, and R-SkL~IV at $(10,1.00)$, primitive translations can be chosen as $(6,0)$ and $(0,6)$, in units of the lattice constant.
Each primitive magnetic unit cell therefore contains 36 sites, so that $n_{\mathrm{mag}}=1$.
For S-SkL~II at $(\Delta,h)=(30,4.00)$ and R-SkL~III at $(30,2.85)$, the primitive translations $(3,3)$ and $(-3,3)$ enclose 18 sites, giving $n_{\mathrm{mag}}=2$.
The corresponding skyrmion-number magnitudes per primitive magnetic unit cell are $|N_{\mathrm{sk}}|=1$ for S-SkL~I, S-SkL~II, R-SkL~II, and R-SkL~III, and $|N_{\mathrm{sk}}|=2$ for R-SkL~IV.

R-SkL~III and R-SkL~IV thus have the same total topological-charge magnitude $|N_{\mathrm{sk}}^{\mathrm{tot}}|=2$, but different primitive magnetic unit cells and skyrmion numbers per cell.
For R-SkL~III, the computational cell contains two primitive magnetic cells, each carrying $|N_{\mathrm{sk}}|=1$; for R-SkL~IV, it contains one primitive magnetic cell carrying $|N_{\mathrm{sk}}|=2$.
Their Fourier contents also differ: the axial components at $\mathbf Q_5$ and $\mathbf Q_6$ vanish in R-SkL~III, whereas R-SkL~IV has unequal axial intensities.
Their appearance demonstrates that spontaneous fourfold-symmetry breaking can occur in a variety of topologically and spatially distinct SkLs.
The two-orbital model therefore supports a broader variety of rectangular SkLs without introducing an explicit in-plane symmetry-breaking interaction.
Instead, the reduction of the fourfold symmetry occurs spontaneously through the redistribution of Fourier components among symmetry-related wave vectors.
This variety highlights the role of the additional crystal-field degrees of freedom in enriching the competition among multiple-$Q$ states.

\subsubsection{Single-$Q$ states}

\begin{figure*}[tp]
\centering
\includegraphics[
width=\textwidth,
height=0.78\textheight,
keepaspectratio
]{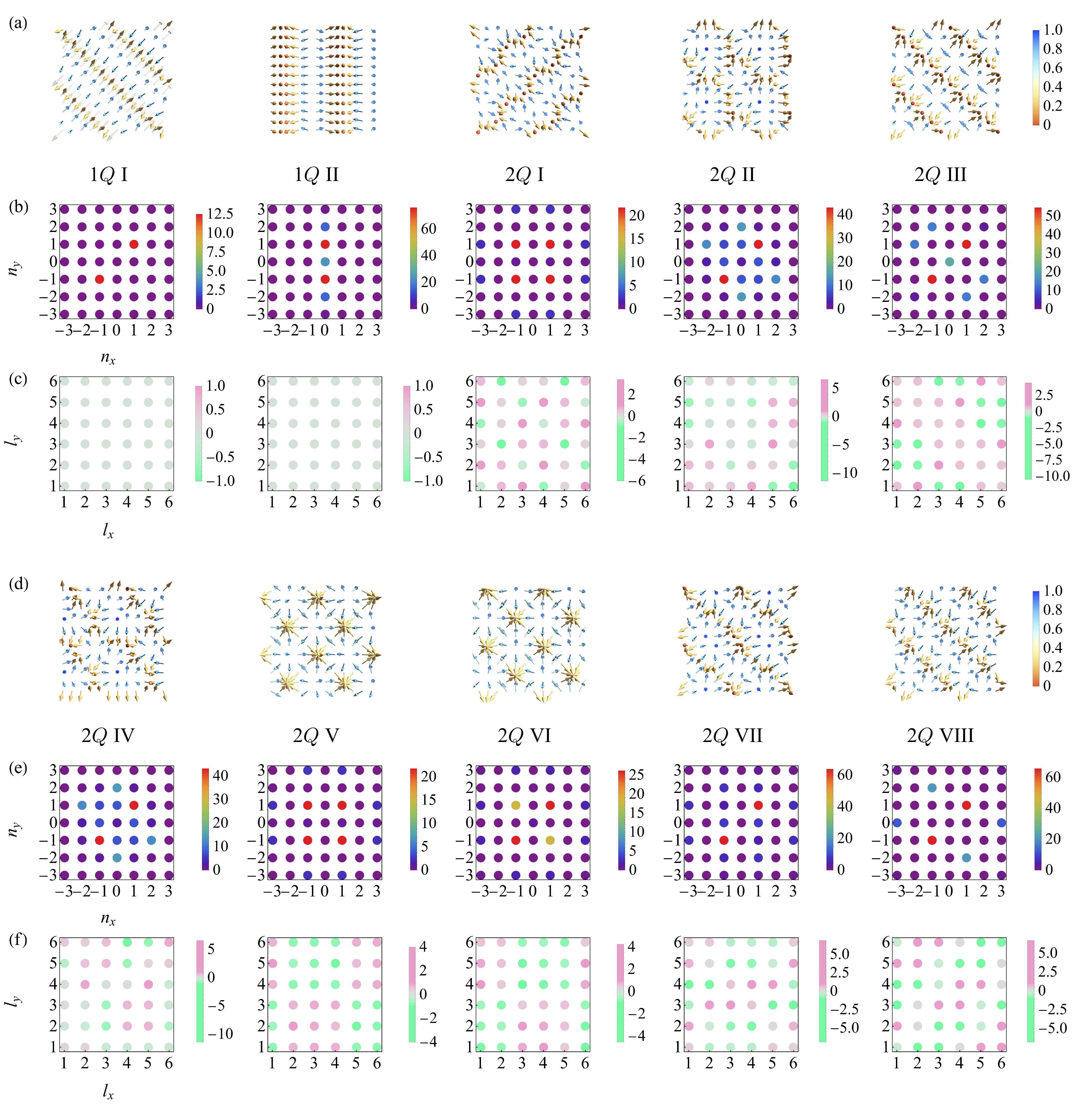}
\caption{
Representative single-$Q$ and double-$Q$ magnetic configurations.
The upper group shows, from left to right, the single-$Q$ vertical state ($1Q$~I), the axial single-$Q$ state ($1Q$~II), $2Q$~I, $2Q$~II, and $2Q$~III, while the lower group shows $2Q$~IV, $2Q$~V, $2Q$~VI, $2Q$~VII, and $2Q$~VIII.
Rows (a) and (d) display the real-space mean moments, with color representing the normalized polar angle $\theta/\pi$; rows (b) and (e) show the total magnetic structure factor $J(\mathbf q)$, where $\mathbf q=Q(n_x,n_y)$ and $Q=\pi/3$; and rows (c) and (f) show the local scalar chirality $\chi_i$.
Panel (a) displays a $2\times2$ periodic tiling of the $6\times6$ computational cell, while panel (c) shows one cell.
The upper and right boundary points in panel (b) are periodic copies of those on the opposite boundaries.
The structure-factor and chirality panels use separate color scales.
The magnetic-phase classification and the magnitudes of the
total topological charges are summarized in
Table~\ref{tab:magnetic_phase_classification}.
The triangulation dependence of the $2Q$~III state is explained
in the table footnote.
}
\label{fig:two_orbital_non_skyrmion}
\end{figure*}

The $1Q$~II state differs from the diagonal $1Q$~I state by its modulation direction.
Whereas $1Q$~I is dominated by $\mathbf Q_1$, the representative $1Q$~II configuration in Fig.~\ref{fig:two_orbital_non_skyrmion} contains the axial fundamental component $\mathbf Q_6$ and its second harmonic $\mathbf Q_4=2\mathbf Q_6$.
These components describe a modulation along the same axial direction.
A symmetry-related domain instead has the corresponding components at $\mathbf Q_5$ and $\mathbf Q_3=2\mathbf Q_5$, as listed in Table~\ref{tab:magnetic_phase_classification}.
It is therefore still a single-$Q$ state rather than a genuine double-$Q$ state.
The emergence of $1Q$~II shows that the participation of the second doublet can rotate the preferred modulation from a diagonal to an axial direction even though the bare exchange--DMI parameters are unchanged.

\subsubsection{Double-$Q$ states}

The two-orbital model also supports several double-$Q$ states.
They are distinguished by the relative weights of the diagonal, combination, and axial Fourier components, as summarized in Table~\ref{tab:magnetic_phase_classification}.
The appearance of several distinct double-$Q$ states indicates that the additional crystal-field degree of freedom substantially enriches the competition among different modulation channels.
In particular, changes in $\Delta$ redistribute the spectral weight among these wave vectors, leading to different multiple-$Q$ textures.

Several double-$Q$ states exhibit finite local scalar chirality
while retaining zero total topological charge, as listed in
Table~\ref{tab:magnetic_phase_classification}.
For the $2Q$~III state, the geometrical topological charge depends on the plaquette triangulation. 
We list zero in Table~\ref{tab:magnetic_phase_classification} and note this ambiguity in the footnote. 

The double-$Q$ labels characterize the modulation content of
the magnetic textures.
Thus, neither multiple-$Q$ character nor local noncoplanarity
alone is sufficient to identify a SkL; the topological charge
together with the real-space texture is essential for
distinguishing the topological and nontopological multiple-$Q$
states.

At fixed $J_1$, $J_2$, and $D$, varying $\Delta$ does not modify the bare exchange--DMI interaction.
The substantial changes in ordering direction, Fourier content, and topological charge therefore originate from the local crystal-field degrees of freedom and their interorbital matrix elements.
This provides a clear distinction between two mechanisms for modifying SkLs: changing the momentum dependence of the interaction, as in the $J_2$ extension, and changing the local Hilbert space at fixed interaction kernel, as in the finite-$\Delta$ calculations.
The latter is specific to the multiorbital $4f$-electron description and has no direct counterpart in a conventional fixed-length spin model.

\section{Conclusion}

\label{sec:conclusion}

We have investigated square and rectangular SkLs in a noncentrosymmetric localized-moment model on a square lattice using self-consistent mean-field calculations.
We first considered a single crystal-field doublet with nearest-neighbor exchange and polar-type DM interaction.
Within the parameter range and computational cell considered, the model supports two square SkLs, S-SkL~I and S-SkL~II, with distinct magnetic periods, as well as R-SkL~I, which spontaneously breaks the in-plane fourfold rotational symmetry.
Their stability changes markedly with the crystal-field mixing coefficient $\alpha$, leading to different field-induced phase sequences and reentrant behavior of R-SkL~I.

These findings demonstrate that the crystal-field wave function provides a microscopic mechanism for controlling both the stability and spatial symmetry of SkLs without explicitly breaking the fourfold symmetry of the Hamiltonian.

We further examined the effects of second-neighbor exchange and of the second crystal-field doublet.
Along the parameter path where the DMI strength is adjusted to keep the diagonal reference wave vectors stationary, $J_2$ substantially modifies the competition among the square and rectangular SkLs, with its effect depending strongly on $\alpha$.
Retaining the second crystal-field doublet produces an even richer phase structure.
As the crystal-field splitting $\Delta$ increases, the phase sequence and phase boundaries progressively approach those of the single-orbital model, while the detailed magnetic texture can converge more slowly.
At smaller $\Delta$, additional rectangular SkLs with distinct
Fourier contents and magnetic periodicities appear together
with axial single-$Q$ states and
various double-$Q$ states.
In particular, R-SkL~I and R-SkL~II have the same total topological-charge magnitude $|N_{\mathrm{sk}}^{\mathrm{tot}}|=1$ but exhibit different distributions of the Fourier components responsible for the broken fourfold symmetry, showing that the topological charge alone does not fully characterize the magnetic texture.
Likewise, R-SkL~III and R-SkL~IV both have $|N_{\mathrm{sk}}^{\mathrm{tot}}|=2$, while their primitive magnetic unit cells carry $|N_{\mathrm{sk}}|=1$ and 2, respectively, demonstrating the importance of distinguishing the total charge of the computational cell from the skyrmion number per primitive magnetic unit cell.

Overall, our results establish the crystal-field degrees of freedom as an important microscopic ingredient in the formation and selection of square and rectangular SkLs in localized-moment systems.
The crystal-field wave function simultaneously controls the effective magnetic anisotropy and coupling strengths, while the participation of excited crystal-field states provides additional degrees of freedom for reorganizing the Fourier content of multiple-$Q$ states.
Together with further-neighbor exchange, these effects generate a variety of topologically and spatially distinct magnetic textures without requiring an explicit in-plane symmetry-breaking interaction.
Our results therefore highlight the importance of treating the local crystal-field structure explicitly when describing multiple-$Q$ magnetism and assessing the validity of reduced single-orbital models.

\begin{acknowledgments}
Cong Guan would like to express her gratitude to Yixiang Liu and ChengYu Chan from Hokkaido University for their helpful discussions.
This research was supported by JSPS KAKENHI Grants Numbers JP22H00101, JP23H04869, JP26K22277, and by JST CREST (JPMJCR23O4 and JPMJCR2641) and JST FOREST (JPMJFR2366). 
Y.S.Z. was supported by JST SPRING (Grant Number JPMJSP2119).
We used OpenAI's GPT-6.1 model to assist with LaTeX typesetting and with checking the code and figures.
\end{acknowledgments}

\appendix

\section{Fourier-space interactions and reference wave vectors}
\label{app:fourier_interaction}

\begin{figure*}[!tp]
\centering
\includegraphics[width=\textwidth]{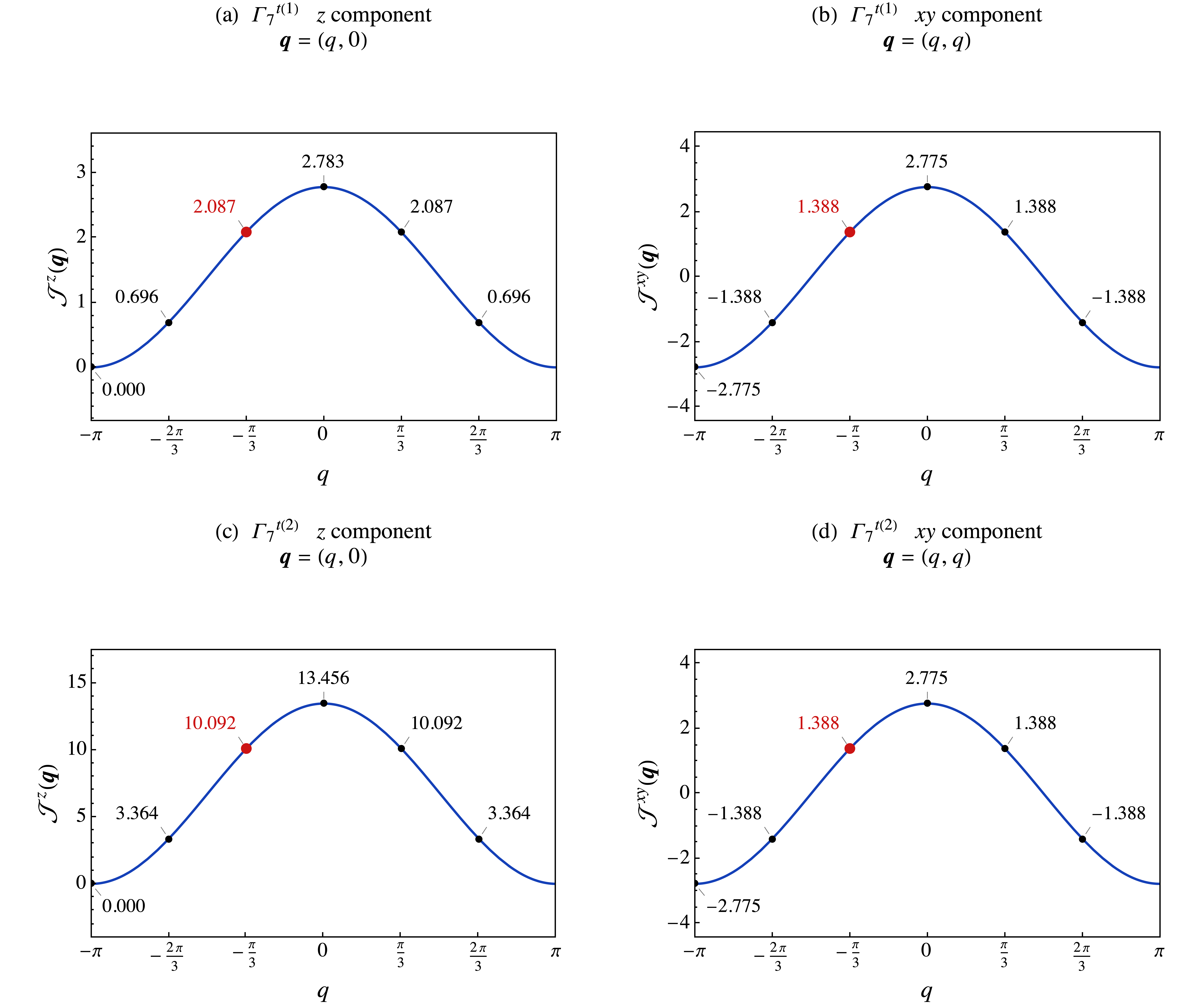}
\caption{
Fourier-transformed interaction functions for the two crystal-field
doublets.
The upper and lower rows correspond to $\Gamma_7^{t(1)}$ and $\Gamma_7^{t(2)}$, respectively, while the left and right columns show the $z$ and $xy$ components of the interaction, respectively.
The momentum dependence is shown along the axial direction $\mathbf q=(q,0)$ in the left column and the diagonal direction $\mathbf q=(q,q)$ in the right column.
The blue curves represent the interaction functions, with the marked points indicating the sampled momenta.
The red points highlight the reference momentum $q=-\pi/3$.
}
\label{fig:app_fourier_interaction}
\end{figure*}

The Fourier transform of the exchange and DMI terms provides a momentum-space representation of the interactions.
Figure~\ref{fig:app_fourier_interaction} shows representative Fourier-transformed interaction functions for the two crystal-field doublets.
The left panels display the $z$ component along the axial cut $\mathbf q=(q,0)$, whereas the right panels display the $xy$ component along the diagonal cut $\mathbf q=(q,q)$.

The momentum-space interaction function is defined as
\begin{equation}
\mathcal{J}^{\mu}(\mathbf q)
=-\min\operatorname{Eig}\left[
\mathcal H_{\mathrm{ex}}^{(\mu)}(\mathbf q)
+\mathcal H_{\mathrm{DM}}^{(\mu)}(\mathbf q)
\right],
\qquad \mu=xy,z.
\end{equation}

These interaction functions are distinct from the magnetic structure factor $J(\mathbf q)$ defined in Sec.~\ref{sec:method}.

For the $6\times6$ computational cell used in the mean-field calculations, the allowed momenta are
$q_\mu=2\pi n_\mu/6$, where $n_\mu$ is an integer and $\mu=x,y$. The reference value $Q=\pi/3$ is therefore commensurate with the computational cell.
The allowed momenta include the diagonal wave vectors $\mathbf Q_1=(Q,Q)$ and $\mathbf Q_2=(Q,-Q)$, the combination wave vectors $\mathbf Q_3=(2Q,0)$ and $\mathbf Q_4=(0,2Q)$, and the axial
wave vectors $\mathbf Q_5=(Q,0)$ and $\mathbf Q_6=(0,Q)$ introduced in the main text.

The continuous curves in Fig.~\ref{fig:app_fourier_interaction} show the momentum dependence of the interaction, whereas the mean-field calculation accesses only the discrete momenta compatible with the
computational cell.
Consequently, an extremum of the continuous interaction function between two allowed momenta does not necessarily coincide with the wave vector selected in the finite-cell calculation.
Furthermore, the axial and diagonal cuts probe the interaction only along the specified directions and should not be interpreted as a search over the full two-dimensional Brillouin zone.
The reference wave vectors $\mathbf Q_1$--$\mathbf Q_6$ therefore provide a convenient basis for analyzing the Fourier content of the self-consistent states, including the axial modes that emerge in the two-orbital calculations.

Finally, the separate-doublet plots illustrate how the crystal-field wave functions modify the momentum-space interactions associated with each doublet.
For finite crystal-field splitting, however, the two-orbital mean-field calculation is performed using the full local angular-momentum matrices, including the interorbital matrix elements.
The resulting magnetic state is therefore determined by the self-consistent response of the full local Hilbert space and cannot be inferred by simply superposing the interaction functions of the two individual doublets.

\section{Parameter path for second-neighbor exchange}
\label{app:J2_parameter_path}

\begin{figure}[!tp]
\centering
\includegraphics[width=\columnwidth]{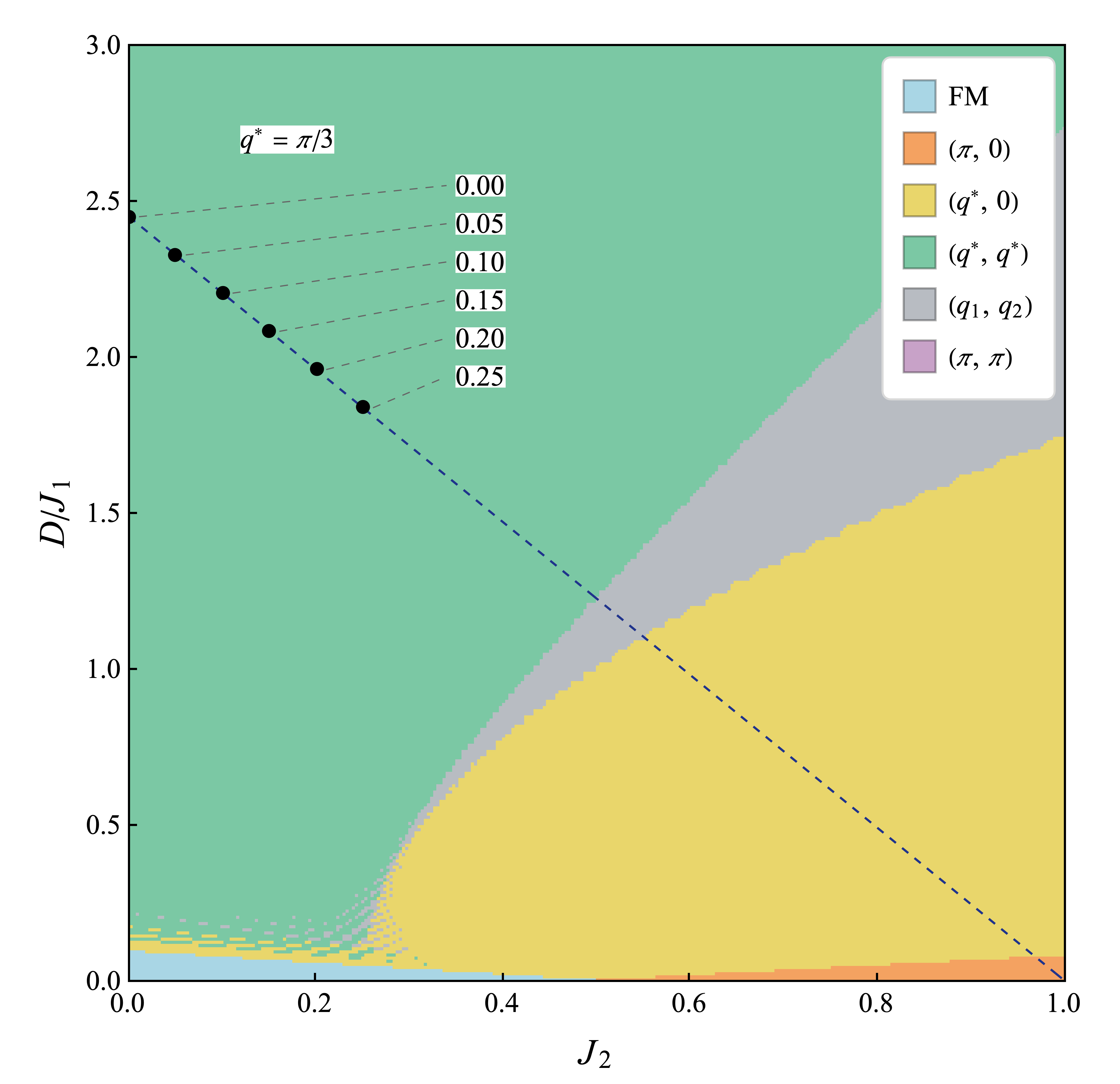}
\caption{
Wave-vector selection of the bare exchange--DMI interaction for negative $J_2$.
The axes represent $|J_2|/J_1$ and $D/J_1$, with the positive-DMI convention adopted in the figure.
The colors indicate the classes of selected wave vectors shown in the legend; FM denotes $\mathbf q=\mathbf 0$.
The blue dashed line represents the parameter path $D/J_1=\sqrt{6}(1-|J_2|/J_1)$, along which the diagonal reference wave vector has $q^*=\pi/3$, where $q^*$ denotes the magnitude of the selected 
wave-vector component. 
The black points indicate representative parameter sets on this path, with the labels specifying $|J_2|/J_1$.
This wave-vector classification refers solely to the bare interaction and should not be confused with the magnetic-phase classification obtained from the self-consistent mean-field calculations.
}
\label{fig:app_J2_parameter_path}
\end{figure}

Figure~\ref{fig:app_J2_parameter_path} illustrates the parameter path used to examine the effect of second-neighbor exchange in momentum space.
The colored regions indicate the wave-vector class favored by the bare exchange--DMI interaction as a function of $|J_2|/J_1$ and $D/J_1$.
This classification characterizes the bare interaction and should be distinguished from the magnetic phases obtained from the self-consistent mean-field calculations in Sec.~\ref{results:second_neighbor}.

As described in the Model section, we vary $J_2$ while setting $J_1=1$ and adjusting the DMI strength according to 
\begin{equation}
 |D|=\sqrt{6}(J_1+J_2).
\end{equation}
This choice keeps the diagonal reference wave vectors $\mathbf Q_1=(Q,Q)$ and $\mathbf Q_2=(Q,-Q)$ with $Q=\pi/3$ as stationary points of the bare interaction. 
For $J_2<0$, the parameter path can be written as
\begin{equation}
 \frac{|D|}{J_1}
 =
 \sqrt{6}\left(1-\frac{|J_2|}{J_1}\right),
\end{equation}
corresponding to the descending dashed line in Fig.~\ref{fig:app_J2_parameter_path}.
The black points mark representative parameter sets sampled along this path and remain within the diagonal wave-vector region shown in the figure.
Only the negative-$J_2$ part of the path is displayed; for positive $J_2$, the calculations follow the same prescription, with $|D|$ increasing with $J_2$.

Although this construction retains the stationary diagonal reference wave vectors, it does not keep the interaction strengths at other momenta fixed.
In particular, the relative weights of the interactions at the combination and axial wave vectors change along the path.
At the same time, the DMI strength and the overall interaction scale also vary with $J_2$.
The evolution of the magnetic phases should therefore be understood as the response along this controlled interaction path, rather than as the effect of $J_2$ alone.
This momentum-space perspective provides the interaction-level context for the different responses found at $\alpha=0.45$ and $0.50$ in the mean-field phase diagrams.

\bibliography{ref.bib}
\bibliographystyle{apsrev4-2}

\end{document}